\documentclass[useAMS,usenatbib]{mn2e}
\usepackage{graphicx}
\usepackage{natbib}

\def\aap{A\&A}
\def\apj{ApJ}

\def\mnras{MNRAS}
\def\pasj{PASJ}
\def\nat{Nat}

\renewcommand{\vec}[1]{\mbox{\boldmath$#1$}}
\newcommand{\sn}{{\rm sn}}
\newcommand{\D}{\displaystyle}
\newcommand{\DF}[2]{\frac{\D#1}{\D#2}}

\begin{document}

\title[Line profiles from thick discs]
{Iron line profiles and self-shadowing from relativistic thick
accretion discs}
\author[Wu and Wang]
{Sheng-Miao Wu\thanks{E-mail:shengmwu@mail.ustc.edu.cn} and Ting-Gui Wang
\thanks{E-mail:twang@ustc.edu.cn }\\
Centre for Astrophysics, University of Science and Technology of
China, Hefei, 230026, China}

\maketitle

\date{Accepted . Received ; in original form}

\markboth{Wu & Wang: Line profiles from thick
discs}{}

\begin{abstract}
We present Fe K$\alpha$ line profiles from and images of relativistic 
discs with finite thickness around a rotating black hole using a novel 
code. The line is thought to be produced by iron fluorescence of a 
relatively cold X-ray illuminated material in the innermost parts of the 
accretion disc and provides an excellent diagnostic of accretion flows 
in the vicinity of black holes. Previous studies have concentrated on 
the case of a thin, Keplerian accretion disc. This disc must become 
thicker and sub-Keplerian with increasing accretion rates. These can 
affect the line profiles and in turn can influence the estimation of the 
accretion disc and black hole parameters from the observed line profiles. 
We here embark on, for the first time, a fully relativistic computation 
which offers key insights into the effects of geometrical thickness and 
the sub-Keplerian orbital velocity on the line profiles. We include all 
relativistic effects such as frame-dragging, Doppler boost, time dilation, 
gravitational redshift and light bending. We find that the separation and 
the relative height between the blue and red peaks of the line profile 
diminish as the thickness of the disc increases. This code is also 
well-suited to produce accretion disc images. We calculate the redshift 
and flux images of the accretion disc and find that the observed image of 
the disc strongly depends on the inclination angle. The self-shadowing 
effect appears remarkable for a high inclination angle, and leads to the 
black hole shadow being completely hidden by the disc itself.
\end{abstract}

\begin{keywords}
{accretion, accretion discs --- black hole physics --- galaxies: active 
--- line: profiles --- X-rays: galaxies}
\end{keywords}

\section{Introduction}
\label{intro} The fluorescent K$\alpha$ iron emission line has been
observed in several active galactic nuclei~(AGN) with 
a broad and skewed line profile. The line is thought to be produced
by iron fluorescence of a relatively cold X-ray illuminated material
in the innermost parts of the accretion disc. Bearing in mind that
the line is intrinsically narrow in the local rest-frame of the emitting
material, and is transformed into broad, skewed profile by Doppler 
shifts and gravitational redshift effects, thus the line profile 
encodes the nature of the structure, geometry, and dynamics of 
the accretion flow in the immediate vicinity of the central black 
hole, as well as the geometry of the space-time, thereby providing 
key information on the location and kinematics of the cold material. 
Investigating these spectral features in X-ray luminous black hole 
systems opens a unique window allows us to probe the physics that 
occurs in the vicinity of a black hole, and provides one way to 
test theory of strong field gravity.

Calculations of the line profiles emitted from an accretion disc
around a black hole have been performed by several authors.
Theoretical Fe K$\alpha$ line profiles from a thin disc around 
a Schwarzschild black hole were calculated by \citet{fab89}. 
\citet{lao91} extended those to the extreme Kerr metric. These 
calculations are based on a geometrically thin, optically thick 
accretion disc (hereafter SSD, following \citet{sha73}), on which 
the accreting material is considered to be in Keplerian orbit around 
a central black hole. Further, the line emissivity is assumed to vary 
with $r$ in power-law form. Efforts have been made later on to include 
various physically plausible processes in the accretion flow, such as 
spiral wave, disc warp, and disc thickness \citep{par98,fuk00,har00,har01,
har02,fuk04}, as well as taking into consideration of the geometry and 
the relative motion of the primary X-ray source \citep{rus00a,dab01,lu01,nay01} 
towards a more realistic emissivity distribution. Some authors considered 
also the ionization effect, the emission from plunging region on the 
iron K line and reflection spectrum \citep{mat93,ros93,mat96,rey97}. 

In the calculations, two basic approaches have been used to map the 
disc into the sky plane. The first method follows photon trajectories 
starting from a given initial locus of emission region in the local 
rest frame of the disc to the observer at infinity. In this case 
a transfer function \citep*{cun75,lao91,spe95,wil98} is usually 
introduced as an integration kernel which includes all relativistic 
effects in line profile calculation. The integration for the line flux 
is then performed directly on the surface of the accretion disc. The transfer 
function was first introduced by \citet{cun75}, who presented the numerical 
results for a grid of parameters aiming at estimating the relativistic 
effect on the continuum emission from SSD, and was re-fined and discussed in 
great detail by \citet{spe95}. The second method adopt a ray tracing 
approach \citep*{dab97,fan97,cad98,mul04,cad05}. Following the trajectories 
of photons from the sky plane to the accretion disc, in this method the 
image of the disc on the observer's sky is derived first and then the line 
flux is obtained by integrating over this image, weighted by the redshift 
factor and the radial disc emissivity profile. Recently, \citet{bec04} 
developed a fast, accurate, high-resolution code which can be used to 
generate high-resolution line profiles numerically. \citet{bec05} extended 
it to include the contribution of higher-order photons to the line profiles. 
But all of these approaches are restricted to SSD.

On the other hand, direct imaging of accretion discs around a black
hole is one of the most exciting areas of study to be considered in the 
future. The fact that 
a black hole posses an event horizon makes a black hole cast a shadow 
upon the background light with a size of roughly ten gravitational radii 
that is due to the bending of light by the black hole, and this shadow 
is nearly independent of the spin or orientation \citep{fal00,zak05}. 
However, for a black hole embedded in an optically thick accretion flow 
the shape and position of the shadow will be altered regardless of the 
black hole spin \citep{tak04,wat05}. From an observational point of 
view, the highest angular resolution is obtained with Very Long Baseline 
Interferometry at millimetre wavelengths (so called mm-VLBI) with an
angular resolution of a few ten micro-arcseconds. This corresponds
to a spatial resolution of only a few ten gravitational radii for
the nearby galaxies. Future Global mm-VLBI Array at short millimetre
wavelengths therefore should allow to map the direct vicinity of the
Super Massive Black Holes (SMBH) such as Sgr A* and M87, and offers
new possibilities to study the immediate environment of SMBH 
\citep{kri04,shen05,bro06,yuan06}. In the X-ray band, the proposed 
Micro-Arcsecond X-ray Interferometry Mission (MAXIM) aims to obtain 
submicroarcsecond resolution X-ray images of nearby galactic nuclei 
\citep{rey03,cas05}. At this resolution, one can capture the image of 
an event horizon of the central massive black hole in a nearby AGN. 
The combination of high resolution radio interferometry and 
interferometric X-ray spectroscopy would form an powerful tool to study 
SMBH and their environment with high accuracy and provide unprecedented 
and direct constraints on the dynamics and geometry of the disc, as well 
as the geometry of the space-time.

With the development of the observational techniques, high-quality
observational data will eventually become available. By fitting the data,
one can in principle constrain the parameters of the accretion disc 
system, this will provide both a direct evidence for the existence 
of a black hole and a way of quantitative test general relativity in 
strong gravity. However, accurate quantitative analysis of observational 
data requires a sophisticated model that treats all relativistic effects 
with a realistic accretion disc structure. At present, such a complete 
model is still not available. To our knowledge, SSD breaks down when the 
accretion rate approaches the Eddington rate. At this limit the disc must 
become geometrically thick and be sub-Keplerain \citep{abr88,wan99,sha05}, 
that is the so called slim disc. For a thick disc, \citet{nan95} pointed 
out it would be of lower density than a standard $\alpha-$disc which would 
increase the ionization parameter~(flux divided by density), thus leads to 
iron in the inner parts of the disc becoming fully ionized and no iron lines 
at all. For slim disc, this may not be the case, the broad, ionized Fe 
K$\alpha$ line was discovered in some narrow-line Seyfert 1 galaxies (NLS1) 
\citep*{bal01,bol02,bol03,fab04,gal07}, which have been thought to work with 
high accretion rates. So, slim disc has been received much more attention 
because of it can be basically used to account for spectral features 
in NLS1 \citep{min00,wan03,che04}. With the increasing evidence for ionized 
accretion discs in NLS1, the spectra and emission lines of slim discs need 
to be studied in more details. Motivated by the above considerations, a 
geometrically and optically thick accretion disc model is presented making 
an attempt at gaining an insight into the effects of disc geometry and 
dynamics on the line profiles and disc images. Following the idea presented 
by \citet{spe95}, we extend their method to the finite thick disc, and adopt 
elliptic integrals to improve the performance of the code which is much 
faster than the direct integral and widely used by many authors.

The paper is organized as follows. In \S\ref{method} we summarize the 
assumptions behind our model, and present the basic equations relevant to 
our problem, while some more technical aspects like the formulae of the 
integration for photon trajectories expressed in terms of the inverse Jacobian 
elliptic functions are given in Appendix~\ref{integral}. We present our results 
in \S\ref{result}, and summarize the conclusions and discussion in \S\ref{sum}.

\section{Assumptions and Method of Calculation}
\label{method}

The aim of this paper is to consider how the accretion disc geometry
and dynamics affect the iron K$\alpha$ line profiles and disc
images. To this end, the disc shape and structure must be
determined first. To obtain a rigorous model, one should solve 
the disc structure equations numerically. However, this is beyond 
the scope of the current work. For simplicity, we adopt a conical 
surface for the disc geometry. The thickness of the disc can be described 
by the half subtending angle $\delta$ ($0 \le \delta \le \pi/4$). When 
$\delta = 0$, the disc reduces to SSD. The complementary angle of
$\delta$ is denoted by $\vartheta_{\rm e}$ which is the angle 
between the symmetric axis of the system and the radial direction of the disc 
surface. The parameters of this model include: the radii of the emitting 
disc zone $r_{\rm in}, r_{\rm out}$, the spin of the black hole $a$, the 
inclination angle of the disc ($\vartheta_{\rm o}$) and the disc surface 
angle ($\vartheta_{\rm e}$), the radial emissivity index $p$, the angular 
velocity index~(see below) $n$, respectively. In addition to all of those, 
the angular dependence of the emissivity also shall be given.

\subsection{Assumptions and basic equations}
\label{equat}

The propagation of radiation from the disc around a Kerr black hole
and the particle kinematics in the disc were studied by many
authors. We review properties of the Kerr metric and formulae for
its particle orbits, and summarize here the basic equations relevant
to this work. Throughout the paper we use
units in which $G = c = 1$, where $G$ is the Gravitational constant, $c$
the speed of light. The background space-time geometry is
described by Kerr metric. In Boyer-Lindquist coordinates, the Kerr
metric is given by
\begin{eqnarray}
ds^{2} & = & -e^{2\nu}dt^{2}+e^{2\psi}(d\phi-\omega
dt)^2+\frac{\Sigma}{\Delta}dr^{2}+\Sigma d\vartheta^{2} ,
\end{eqnarray}
where
\[e^{2\nu}=\Sigma\Delta/A,\,e^{2\psi}=\sin^{2}\vartheta A/\Sigma,\,\omega=2Mar/A,\]
\[\Sigma=r^{2}+a^{2}\cos^{2}\vartheta,\,\Delta=r^{2}+a^{2}-2Mr,\]
\[A=(r^{2}+a^{2})^{2}-a^{2}\Delta\sin^{2}\vartheta.\] 
Here $M$, $a$ are the black hole mass and specific angular momentum, respectively.

The general orbits of photons in the Kerr geometry can be expressed
by a set of three constants of motion \citep{car68}. Those are the
energy at infinity $E$, the axial component of angular momentum
$E\lambda$, and carter's constant ${\cal Q}\,({=}q^2E^2)$. The
4-momentum of a geodesic has components
\begin{equation}
   p_\mu = (p_{\rm t},\,p_{\rm r},\,p_\vartheta,\,p_\phi) = (-E,\,\pm E\sqrt{R}/
\Delta,\,\pm E\sqrt\Theta,\,E\lambda),
\end{equation}
with
\begin{eqnarray*}
    R &=& r^4 + \left(a^2-\lambda^2-q^2\right)r^2 +2M\left[q^2+(\lambda-a)^2
        \right]r - a^2 q^2 \;,  \\
    \Theta &=& q^2 + a^2 \cos^2\vartheta - \lambda^2 \cot^2\vartheta\;.
\end{eqnarray*}
From this, the equations of motion governing the orbital trajectory
can be obtained. The technical details are given in
Appendix~\ref{integral}.

We assume that the disc is of a cone-shaped surface, axisymmetric, and 
lies on the equatorial plane of the black hole. Photons are emitted or 
reflected from the gas on the conical disc surface which moves along circular 
orbits. The radial drift of the gas on the disc surface is neglected.
Thus, the 4-velocity field is chosen to be of the form
\begin{equation}
  u^\mu = u^t(\partial_{\rm t},\,0,\,0,\,\Omega\partial_\phi) = (u^t,\,0,\,0,\,u^\phi),
\end{equation}
where $\Omega=u^\phi/u^t$ is the angular velocity of the emitting
gas. The choice of $\Omega$ must satisfy the causality condition. For 
sub-Keplerian velocity, we adopt the modification of $\Omega$ firstly 
introduced by \citet{rus00b}
\begin{equation}
\Omega=\left(\frac{\vartheta}{\pi/2}\right)^{1/n}\Omega_K+\left[1-\left(
\frac{\vartheta}{\pi/2}\right)^{1/n}\right]\omega,
\label{omega}
\end{equation}
where $\vartheta$ is the poloidal Boyer-Lindquist coordinate,
$\Omega_K=M^{1/2}/(r^{3/2}+aM^{1/2})$ is the Keplerian angular
velocity and $\omega$ is the angular velocity of the gravitational
drag. It is easy to verify that $\Omega\le\Omega_K.$

For describing physical processes near a Kerr black hole, Boyer-Lindquist 
coordinates, which are unphysical in the ergosphere, are inconvenient. In 
order to make physics appear simple in their frames, the locally nonrotating 
frames~(LNRF) was introduced by \citet*{bar72}. The relation between the 
local rest frame attached to the disc fluid and LNRF is given by a Lorentz 
transformation. In the LNRF, the azimuthal component of 3-velocity reads
\begin{equation}
v=e^{\psi-\nu}(\Omega-\omega)=\frac{A\sin\vartheta}{\Sigma\sqrt\Delta}
(\Omega-\omega).
\end{equation}
The corresponding Lorentz factor $\gamma$ as measured by LNRF is
defined as $\gamma=(1-v^2)^{-1/2}.$

Due to relativistic effects, the photon frequency will shift from 
the emitted frequency $\nu_{\rm e}$ to the observed one $\nu_{\rm o}$ 
received by a rest observer with the hole at infinity. We introduce 
a $g$ factor to describe the shift which is the ratio of observed 
frequency to emitted one:
\begin{eqnarray}
    g & = & \nu_{\rm o} / \nu_{\rm e} = (\vec{p}\cdot\vec{u}_{\rm o})/
    (\vec{p}\cdot\vec{u}_{\rm e}) \nonumber \\
      & = & e^\nu(1 - v^2)^{1/2}/(1 - \Omega\lambda),
     \label{gvalue}
\end{eqnarray}
where $\vec{p}, \vec{u}_{\rm o}, \vec{u}_{\rm e}$ are the 4-momentum
of the photon, the 4-velocity of the observer and the emitter,
respectively.

The specific flux density $F_{\rm o}(\nu_{\rm o})$ at frequency
$\nu_{\rm o}$ as observed by an observer at infinity is defined as
the sum of the observed specific intensities $I_{\rm o}(\nu_{\rm
o})$ from all parts of the accretion disc surface,
\begin{eqnarray}
    F_{\rm o}(\nu_{\rm o}) = \int I_{\rm o}(\nu_{\rm o}) d\Omega_{\rm obs} \;,
    \label{feo}
\end{eqnarray}
where $d\Omega_{\rm obs}$ is the element of the solid angle subtended by 
the image of the disc on the observer's sky. We do not consider the 
effect of higher order images of the disc in the following computations 
as their contribution is small due to most high order photons reintercept 
and be absorbed by the disc.

Using the fact that $I(\nu)/\nu^3$ is invariant along the path of a
photon, where $\nu$ is the photon frequency measured by any local
observer on the path , equation~(\ref{feo}) can be rewritten as
\begin{eqnarray}
    F_{\rm o}(\nu_{\rm o}) = \int g^3 I_{\rm e}(\nu_{\rm e}) d\Omega_{\rm obs} \;.
    \label{feo1}
\end{eqnarray}
$I_{\rm e}(\nu_{\rm e})$ is the specific intensity measured by an
observer corotating with the disc, and can be approximated by a $\delta$-function, 
$I_{\rm e}^{\prime}(\nu_{\rm e}^{\prime})=\varepsilon\delta(\nu_{\rm
e}^{\prime}-\nu_{\rm e})$ where $\varepsilon$ is the emissivity per unit 
surface area. From well known transformation properties of
$\delta$-functions we have $\delta(\nu_{\rm e}^{\prime}-\nu_{\rm
e})=g\delta(\nu_{\rm o}-g\nu_{\rm e})$, using this in
equation~(\ref{feo1}), we obtain
\begin{eqnarray}
    F_{\rm o}(\nu_{\rm o}) = \int \varepsilon g^4 \delta(\nu_{\rm o}-g\nu_{\rm e}) d\Omega_{\rm obs}\;.
    \label{feo2}
\end{eqnarray}

In order to calculate the integration over $d\Omega_{\rm obs}$, we
must first obtain the disc image or find the relation between the
element of the solid angle and the disc linked by the null geodesic.
The apparent position of the disc image as seen by an observer is
conveniently represented by two impact parameters $\alpha$ and
$\beta$, measured relative to the direction to the centre of the
black hole. The impact parameters $\alpha$ and $\beta$ are,
respectively, the displacements of the image in the directions of 
perpendicular and parallel to the projection of the black hole spin. 
They are related to two constants of motion $\lambda$ and $q$ by
\citep{cun73,cun75}
\begin{eqnarray}
    \alpha = - \lambda/\sin\vartheta_{\rm o},\,
    \beta = \pm\left(q^2 + a^2\cos^2\vartheta_{\rm o} - \lambda^2\cot^2
        \vartheta_{\rm o}\right)^{1/2} \!, \label{alp_beta}
\end{eqnarray}
where $\vartheta_{\rm o}$ is the angle between the observer and the 
rotation axis of the black hole ({\rm i.e.}~the inclination angle). The element 
of solid angle seen by the observer is then
\begin{eqnarray}
    d\Omega_{\rm obs} & = & \frac{d\alpha d\beta}{r_{\rm o}^2} = \frac{1}{r_{\rm o}^2}
     \frac{\partial(\alpha,\beta)}{\partial(\lambda,q)}
     \frac{\partial(\lambda,q)}{\partial(r,g)}\;dr\;dg\nonumber \\
     & = & \frac{q}{r_{\rm o}^2\beta\sin\vartheta_{\rm o}}
     \frac{\partial(\lambda,q)}{\partial(r,g)}\;dr\;dg,
\label{solid_obs}
\end{eqnarray}
where $r_{\rm o}$ is the distance from the observer to the black
hole.

Substituting equation~(\ref{solid_obs}) into equation~(\ref{feo2})
gives the desired result:
\begin{eqnarray}
    F_{\rm o}(\nu_{\rm o}) & = & \frac{q}{r_{\rm o}^2\beta\sin\vartheta_{\rm o}}\int \varepsilon
    g^4 \delta(\nu_{\rm o}-g\nu_{\rm e}) \frac{\partial(\lambda,q)}{\partial(r,g)}\;dr\;dg.
    \label{feo3}
\end{eqnarray}

To perform the integration, the form of the disc emissivity in the
integrand also needs to be given. In general, it can be a function of the
radius, $r_{\rm e}$ and polar angle, $n_{\rm e}$, of an emitted
photon with the surface normal of the disc in the rest frame of the
emitting gas. This angle is determined by taking the dot-products of
the photon four-momentum $\vec{p}$ with the surface normal
$\vec{n}$. The surface normal in the rest frame is
\begin{equation}
    \vec{n} = \Sigma^{-1/2}\partial/\partial\vartheta.
    \label{norm}
\end{equation}
By definition, we get
\begin{eqnarray}
    \cos(n_{\rm e}) & = & \frac{\vec{p}\cdot\vec{n}}{\vec{p}\cdot\vec{u_{\rm e}}}
     = \frac{\vec{p}\cdot\vec{n}}{\vec{p}\cdot\vec{u_{\rm o}}}\frac
     {\vec{p}\cdot\vec{u_{\rm o}}}{\vec{p}\cdot\vec{u_{\rm e}}} \nonumber \\
      & = & g\sqrt{\Theta}/\sqrt{\Sigma} \nonumber \\
      & = & e^\nu(1 - v^2)^{1/2}\Theta^{1/2}\Sigma^{-1/2}/(1 -
      \Omega\lambda).
\end{eqnarray}
If the emission is isotropic in the rest frame, we do not need to
know $n_{\rm e}$. More generally, we take the form
\begin{equation}
   \varepsilon(r_{\rm e},\mu_{\rm e}) = \epsilon(r_{\rm e})f(\mu_{\rm e}),
\label{emis}
\end{equation}
where $\mu_{\rm e}$ is the cosine of the emitted angle~($\cos n_{\rm e}$). 
And the radial emissivity is assumed to vary as a power law with emissivity index $p$:
\begin{equation}
    \epsilon(r_{\rm e}) \propto r_{\rm e}^{-p}.
    \label{emisr}
\end{equation}
We consider three possible cases for the angular dependence of the 
emissivity \citep{bec05}: (1) isotropic emission, $f(\mu_{\rm e})=1$; 
(2) limb darkening law $f(\mu_{\rm e})\propto (1+2.06\mu_{\rm e})$; 
(3) limb brightening law $f(\mu_{\rm e})\propto 1/\mu_{\rm e}$.

\subsection{Method of calculation }
\label{meth}

With all of the preparation described in the previous section, we 
now turn to how to calculate the line profiles and the disc images 
numerically. We divide the disc into a number of arbitrarily narrow 
rings, and emission from each ring is calculated by considering its 
axisymmetry. We shall denote by $r_{\rm i}$ the radius of each such 
emitting ring. For each ring there is a family of null geodesic along 
which radiation flows to a distance observer at polar angle 
$\vartheta_{\rm o}$ from the disc's axis. As far as the iron K$\alpha$ 
emission line is concerned, for a given observed frequency $\nu_{\rm o}$ 
the null geodesic in this family can be picked out if it exists. So, 
the weighted contribution of this ring to the line flux can be determined. 
The total observed flux can be obtained by summing over all emitting rings. 
Changing the observed frequency, then the line profiles will be obtained.

This family of null geodesic for each ring can be used to map the accretion 
disc onto sky plane, that is, disc imaging. A geodesic in this 
family connects an emitting region in the ring to the distant observer. The 
constants of motion $\lambda$ and $q$ of this geodesic can be used to 
determine the apparent position of the emitting point on the sky plane 
using the corresponding two impact parameters $\alpha$ and $\beta$. 
Different geodesic is associated with different point. Using geometric 
optics, one determines the appearance of the ring from this family of geodesic,
then in this way images~(at infinity) of the accretion disc are obtained.

The main numerical procedures for computing the line profiles are as
follows:
\begin{enumerate} 
\item Specify the relevant disc system parameters: $r_{\rm in}, r_{\rm
out}, a, p, n, \vartheta_{\rm o}, \vartheta_{\rm e}$ and the angular 
emissivity. 

\item The disc surface is modeled as a series of rings with the
radii $r_{\rm i}$ and weights $\omega_{\rm i}$ which calculated
using an algorithm due to Rybicki G. B. \citep{pre92}.

\item For a given couple~($r_{\rm i},g$) of a ring, the two constants of 
motion $\lambda$ and $q$ are determined if they exist. This is done in the 
following way: the value of $\lambda$ is obtained by the another form of 
equation~(\ref{gvalue}) 
\begin{eqnarray}
    \lambda & = & \frac{1}{\Omega}\left(1-\frac{e^\nu}{\gamma g}\right)
     =\frac{1}{\Omega}\left(1-\frac{e^\nu(1 -
    v^2)^{1/2}}{g}\right),
    \label{lambda}
\end{eqnarray}
the value of $q$ is determined for solving photon trajectory 
equation~(\ref{r-th}). Then the contribution of this ring on the flux 
for given frequency $\nu_{\rm o}$ with respect to $g$ is estimated.

\item For a given g, the integration over $r$ of the equation~(\ref{feo3}) 
can be replaced by a sum over all the emitting rings
\begin{eqnarray}
    F_{\rm o}(\nu_{\rm o}) & = & \sum_{i=1}^n \frac{q \varepsilon \nu_{\rm o}^4}
    {r_{\rm o}^2 \nu_{\rm e}^4 \beta\sin\vartheta_{\rm o}}
     \left.\frac{\partial(\lambda,q)}{\partial(r,g)}\right|_{\rm r=r_i}\omega_{\rm i}.
    \label{feo4}
\end{eqnarray}
The Jacobian $[\partial(\lambda,q)/\partial(r,g)]$ in the above formula 
was evaluated by the finite difference scheme. From above formula, one 
determines the line flux at frequency $\nu_{\rm o}$ from the disc.

\item Varying $g$, the above steps are repeated.
\end{enumerate}
 
The observed line profile as a function of frequency $\nu_{\rm o}$
is finally obtained in this way.

\section{Results}
\label{result}

The model and the computer code described in this paper suitable 
for disc inner edge located at any $r_{\rm in}\ge r_{\rm ms}$, 
where $r_{\rm ms}$ is the radius of the marginally stable orbit. 
For simplicity, in all plots presented in the paper, we assume that 
$r_{\rm in} = r_{\rm ms}$. We have taken a disc from $r_{\rm ms}$ 
to $r_{\rm max}=20r_{\rm g}$~(focus on strong gravitational effects) 
for Kerr metric case treated spin of the black hole as a free parameter 
for different observed inclinations and disc thickness. Due to its 
astrophysical importance, we choose the iron fluorescence line at 
$6.4\,{\rm keV}$ in what follows.

\subsection{Relativistic emission line profiles}
\begin{figure}
\includegraphics[width=\hsize]{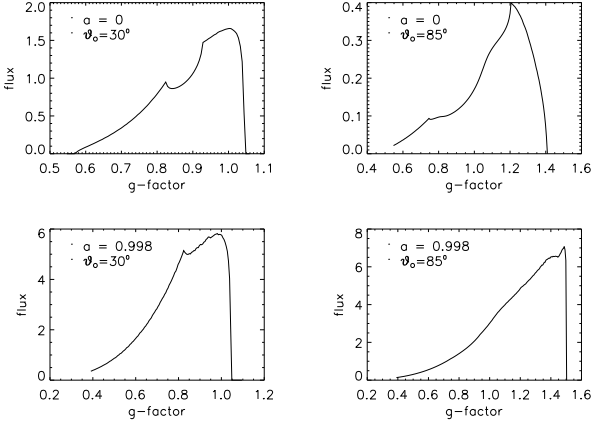}
  \caption{The relativistic line profiles computed by our code both for the
 Schwarzschild~($a=0$) and maximal Kerr~($a=0.998$) cases for $\vartheta_{\rm o}
  = 30^{\circ}$~(left) and $85^{\circ}$~(right). The disc zone is from $r_{\rm ms}$ to
  $r_{\rm out}= 20 r_{\rm g}$ and located at equatorial plane, where $r_{\rm g}$ is
  the gravitational radius. Upper panel: The Schwarzschild metric for $\epsilon(r_{\rm
 e}) \propto r_{\rm e}^{-3}$ and $f(\mu_{\rm e})=1$. Lower panel: The
 maximal Kerr metric for $\epsilon(r_{\rm e}) \propto r_{\rm e}^{-3}$
 and $f(\mu_{\rm e}) \propto (1+2.06 \mu_{\rm e})$. The flux in all
 cases is given using the same arbitrary units, and all our results
 are unsmoothed.
 \label{lineprof}}
\end{figure}

\begin{figure}
\includegraphics[width=\hsize]{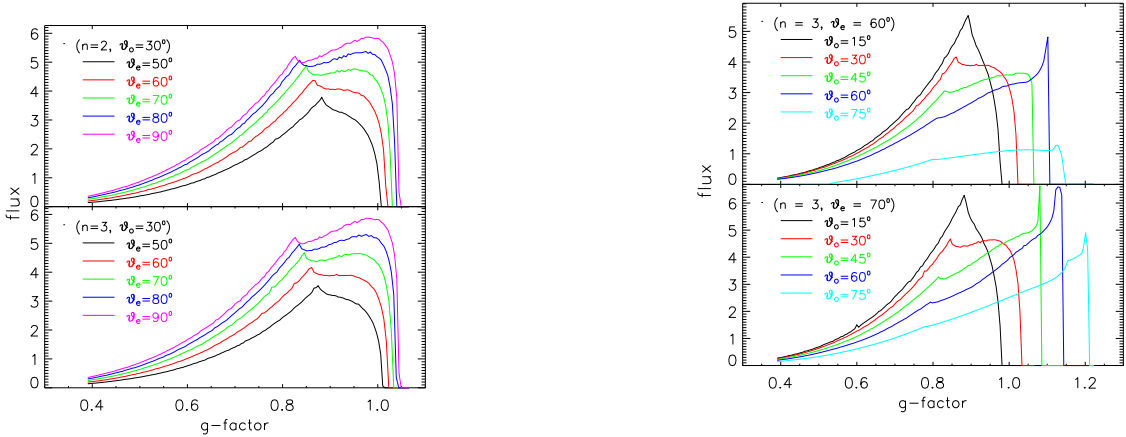}
\caption{The relativistic line profiles as a function of the disc
thickness for (from bottom to top at the red peak) $\vartheta_{\rm
e} = 50^{\circ},60^{\circ},70^{\circ},80^{\circ} \,\rm and \;
90^{\circ}$ for a maximal Kerr black hole with the disc extending
from $1.235 - 20 r_{\rm g}$. The observer inclination equals
$30^{\circ}$ and angular velocity takes the form:
$\Omega=\left(\frac{\vartheta}{\pi/2}\right)^{^{1/n}}\Omega_K+\left[1-\left(
\frac{\vartheta}{\pi/2}\right)^{^{1/n}}\right]\omega$, here n is set
to 2~(upper panel) and 3~(lower panel). The emissivity law is taken
the forms $\epsilon(r_{\rm e}) \propto r_{\rm e}^{-3}$ and
$f(\mu_{\rm e}) \propto (1+2.06 \mu_{\rm e})$.
 \label{thickf}}
\end{figure}

\begin{figure}
\includegraphics[width=\hsize]{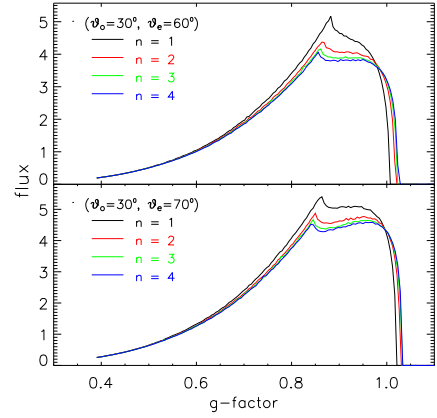}
\caption{The relativistic line profiles as a function of the angular
velocity represented by the parameter $n$ in equation~(\ref{omega}):
$\Omega=\left(\frac{\vartheta}{\pi/2}\right)^{^{1/n}}\Omega_K+\left[1-\left(
\frac{\vartheta}{\pi/2}\right)^{^{1/n}}\right]\omega\,$  for
$n=1,2,3,4$ (from top to bottom at the redshift peak) for a maximal
Kerr black hole with the disc extending from $1.235 - 20 r_{\rm g}$
and $\vartheta_{\rm o}=30^{\circ},\vartheta_{\rm
e}=60^{\circ}$~(upper panel), $70^{\circ}$~(lower panel). The
emissivity law is the same as in Fig. \ref{thickf}.
 \label{angulv}}
\end{figure}

\begin{figure}
\includegraphics[width=\hsize]{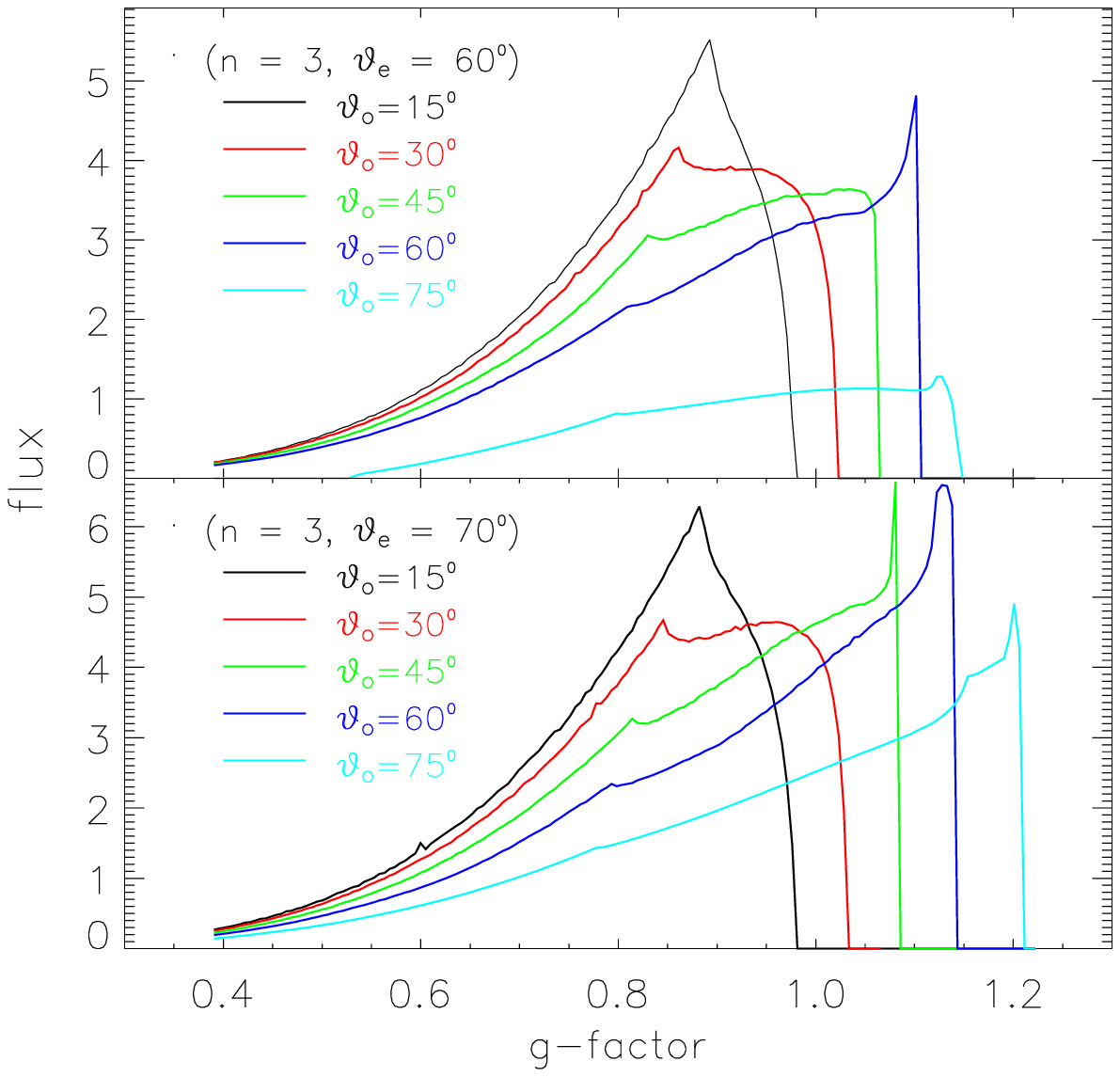}
\caption{The relativistic line profiles as a function of the
observed inclinations for $\vartheta_{\rm o} =
15^{\circ},30^{\circ},45^{\circ},60^{\circ} \,\rm and \, 75^{\circ}$
for a maximal Kerr black hole with the disc extending from $1.235 -
20 r_{\rm g}$ and $\vartheta_{\rm e}=60^{\circ}$~(upper panel),
$\vartheta_{\rm e}=70^{\circ}$~(lower panel). The index $n$ is set to
3 and the emissivity law is the same as in Fig.
\ref{thickf}.
 \label{obsinclin}}
\end{figure}

The numerical code discussed in previous section was used to model 
emission line profiles for different model parameters. To test the 
performance of our code, we first compared the line profiles 
generated by our code when the disc is reduced to SSD to 
those generated by the code described in \citet{bec04}, and  
found that the overall match is fairly good, especially for the 
Schwarzschild metric case. Fig.\ref{lineprof} shows the results for 
parameters identical to those presented by \citet{bec04}.

The dependence of the line profiles on the disc thickness is shown 
in Fig.\ref{thickf}. The angular velocity and the emissivity law 
take the forms:
$\Omega=\left(\frac{\vartheta}{\pi/2}\right)^{^{1/n}}\Omega_K+\left[1-\left(
\frac{\vartheta}{\pi/2}\right)^{^{1/n}}\right]\omega$,
$\epsilon(r_{\rm e}) \propto r_{\rm e}^{-3}$ and $f(\mu_{\rm e})
\propto (1+2.06 \mu_{\rm e})$. Two cases for $n=2$~(top panel) and
$n=3$~(bottom panel) are presented in this figure. It is explicit
that the separation and relative height between the blue and red
peaks diminish as disc thickness increases because the disc becomes more
sub-Keplerian. This effect is also clearly illustrated in
Fig.\ref{angulv}. The index $n$ in $\Omega$ describes the deviation 
of the angular velocity from Keplerian one. As the deviation from 
Keplerian velocity increases the height of the blue peak of the 
line decreases significantly.

Fig.\ref{obsinclin} compares the line profiles at different viewing
angles $\vartheta_{\rm o} = 15^{\circ}, 30^{\circ}, 45^{\circ},
60^{\circ} \,\rm and \, 75^{\circ}$ for a maximal Kerr black hole
with the disc extending from $r_{\rm ms}$ to $20 r_{\rm g}$ and
$\vartheta_{\rm e}=60^{\circ}, 70^{\circ}$. At high inclinations 
the self-shadowing effect has been taken into account. Due to 
gravitational lensing (light bending) effect, there is still
a substantial fraction of light that can reach the observer 
at infinity.

\begin{figure}
\includegraphics[width=\hsize]{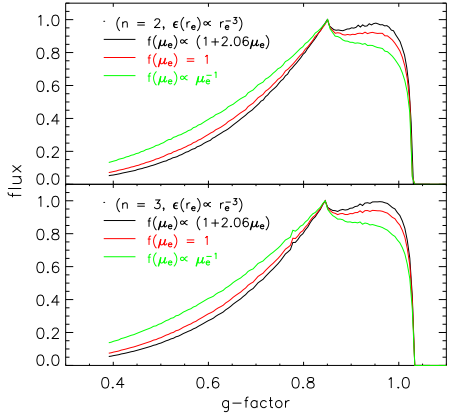}
  \caption{The relativistic line profiles generated by our model with (a)
 $\epsilon \left( r_{\rm e} \right) \propto r_{\rm e}^{-3}$, $f \left(\mu_{\rm e}
 \right) \propto \left( 1 + 2.06 \mu_{\rm e} \right)$ (black line), (b)
 $\epsilon \left( r_{\rm e}\right) \propto r_{\rm e}^{-3}$,
 $f \left( \mu_{\rm e} \right) = 1$ (red line),  (c) $\epsilon \left( r_{\rm e} \right)
 \propto r_{\rm e}^{-3}$, $f \left( \mu_{\rm e} \right) \propto \mu^{-1}_{\rm e}$
 (green line), for a maximal Kerr black hole with the disc extending from
 $1.235 - 20r_{\rm g}$ and $\vartheta_{\rm o}=30^{\circ},\vartheta_{\rm
e}=70^{\circ}$. The sub-Keplerian angular velocity is the same as in
Fig. \ref{thickf}. All profiles are scaled to unity for better
comparison in this case.
 \label{angulem}}
\end{figure}

\begin{figure}
\includegraphics[width=\hsize]{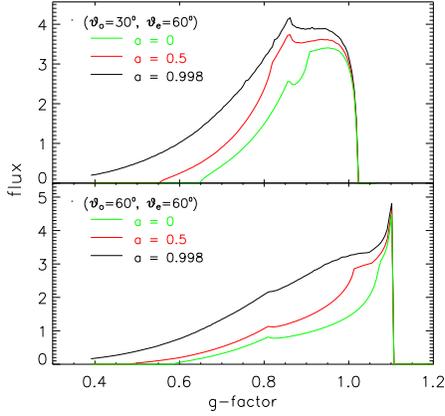}
  \caption{ Comparison of the relativistic line profiles generated by our
model with different spins $a=0,\, 0.5,\, 0.998$. The emission line
region is from $r_{\rm ms}$ to $20r_{\rm g}$ and the angular
velocity and the emissivity law are the same as in Fig. \ref{obsinclin}. 
The angles are marked in each figure.
 \label{spindif1}}
\end{figure}

\begin{figure}
\includegraphics[width=\hsize]{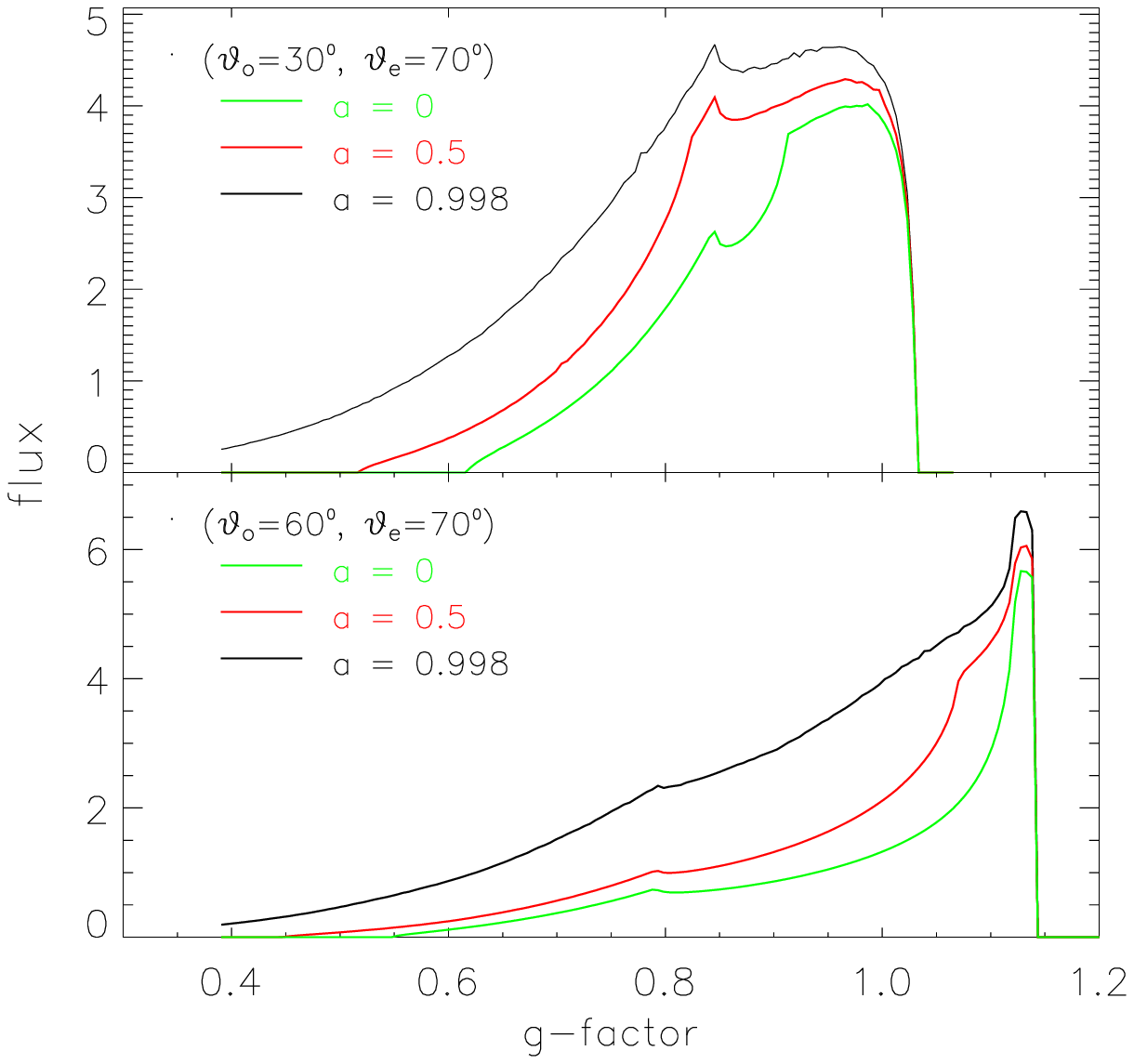}
 \caption{As in Fig. \ref{spindif1} but with the angle $\vartheta_{\rm e}=70^{\circ}$.
 \label{spindif2}}
\end{figure}

\begin{figure}
\includegraphics[width=\hsize]{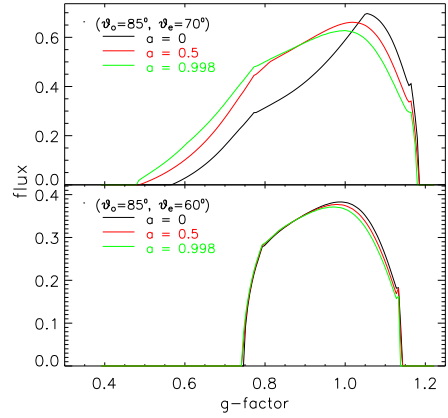}
  \caption{As in Fig. \ref{spindif1} but with the angles $\vartheta_{\rm o} =
85^{\circ}$ and \,$\vartheta_{\rm e}=70^{\circ}$ (upper panel),
$60^{\circ}$ (lower panel).
 \label{spindif3}}
\end{figure}

We also calculated the effects of emissivity on the form of the 
relativistic line profile. The radial emissivity is taken a power 
law with the index $p$, which determines the relative contribution 
from different radii of the disc. Here we focus on the influence of 
anisotropic emission on the line profile. Different angular emissivity 
laws have striking effects on the line profile, which we illustrate 
in Fig.\ref{angulem} for a maximal Kerr 
geometry with the disc extending from $r_{\rm ms}$ to $20 r_{\rm g}$ 
and $\vartheta_{\rm o}=30^{\circ}, \vartheta_{\rm e}=70^{\circ}$. 
The angular emissivity takes one of the three forms: (a)
 $\epsilon \left( r_{\rm e} \right) \propto r_{\rm e}^{-3}$, $f \left(\mu_{\rm e}
 \right) \propto \left( 1 + 2.06 \mu_{\rm e} \right)$, (b)
 $\epsilon \left( r_{\rm e}\right) \propto r_{\rm e}^{-3}$,
 $f \left( \mu_{\rm e} \right) = 1$,  (c) $\epsilon \left( r_{\rm e} \right)
 \propto r_{\rm e}^{-3}$, $f \left(\mu_{\rm e} \right) \propto \mu^{-1}_{\rm
 e}$. 
From the figure one can see the relative height of the blue wing changes 
a lot for different angular emissivity laws, anti-correlated with the 
slope of the red wing.

\begin{figure}
\includegraphics[width=\hsize]{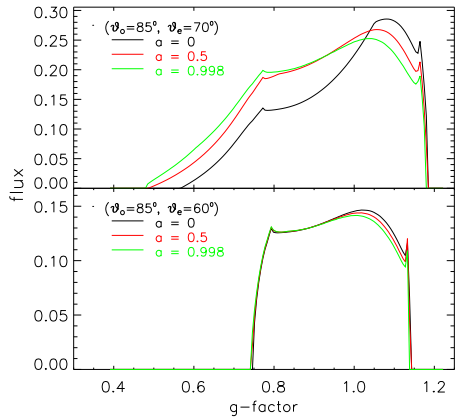}
  \caption{As in Fig. \ref{spindif3} but with no angular dependence
  of emissivity $\epsilon \left( r_{\rm e}\right) \propto r_{\rm e}^{-3}$,
 $f \left( \mu_{\rm e} \right) = 1$.
 \label{spindif4}}
\end{figure}

The line profiles as a function of the black hole spin are also 
demonstrated. For a low or intermediate inclination angle the 
line profiles are shown in Figs \ref{spindif1} and \ref{spindif2}. Note 
that, the red wings change significantly whereas the blue peaks almost are 
not affected by the spin. At high inclinations, the 
effect of the self-shadowing dramatically alter the line profile for 
a thick disc. The results are illustrated in Figs \ref{spindif3} and 
\ref{spindif4} with angular emissivity $f(\mu_{\rm e}) 
\propto (1+2.06 \mu_{\rm e})$ and  $f(\mu_{\rm e}) = 1$, 
respectively. For $\vartheta_{\rm e}=60^{\circ}$, the line profiles 
are almost the same, this implies that the line emission from the inner 
parts of the disc are completely obscured by the outer parts of the disc. 
At high viewing angles, the impact of angular emissivity law on the 
relativistic line profiles are also striking.

\subsection{Accretion disc images}
\begin{figure*}
\includegraphics[width=\hsize]{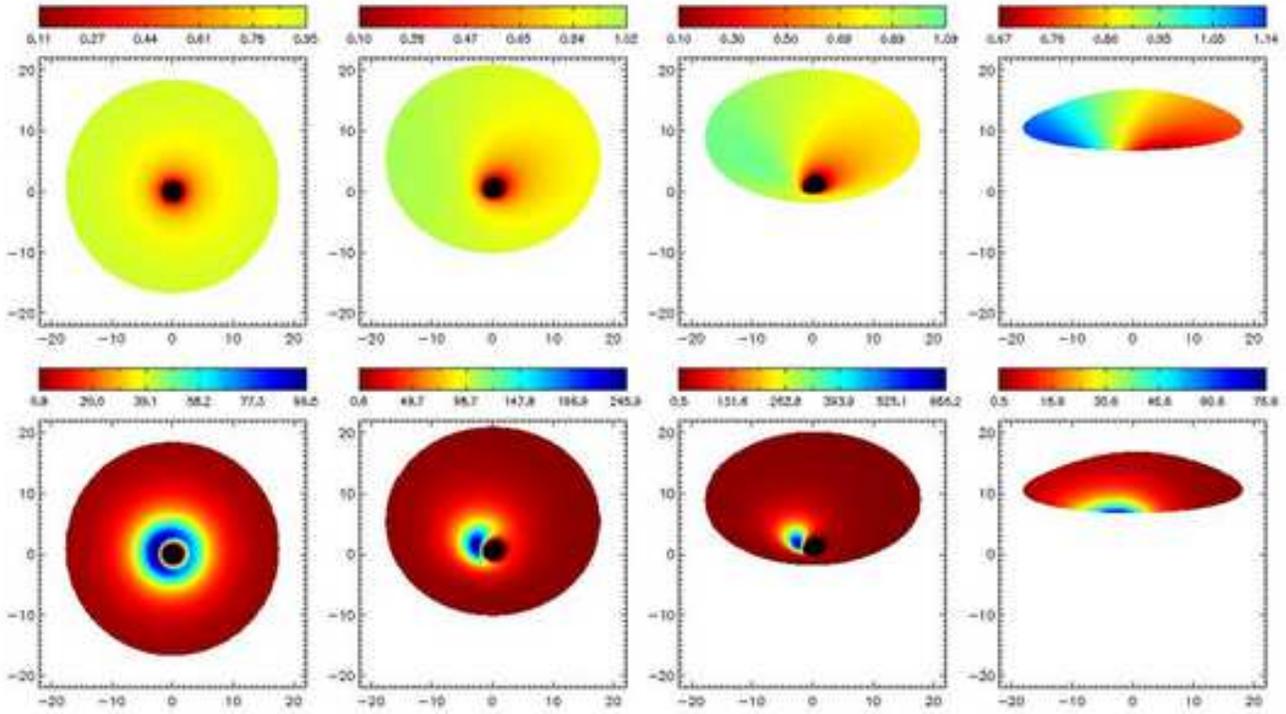}
\vspace{-1.5cm} 
\caption{Redshift images~(top row) and flux images~(bottom row) of the accretion
 disc on the ($\alpha,\beta$) plane for a extreme Kerr black hole. From left- to 
right-hand side: $\vartheta_{\rm o}=5^{\circ}, 30^{\circ}, 55^{\circ}, 80^{\circ}$ 
and $\vartheta_{\rm e}=60^{\circ}$.  Redshift images are colored by the associated 
values of g as measured by the infinity observer. Flux  images are colored by 
$10^4\varepsilon g^4$. The parameters n and p are both set to 3
 and $f(\mu_{\rm e}) \propto (1+2.06 \mu_{\rm e})$.
 \label{image1}}
\end{figure*}

\begin{figure*}
\includegraphics[width=\hsize]{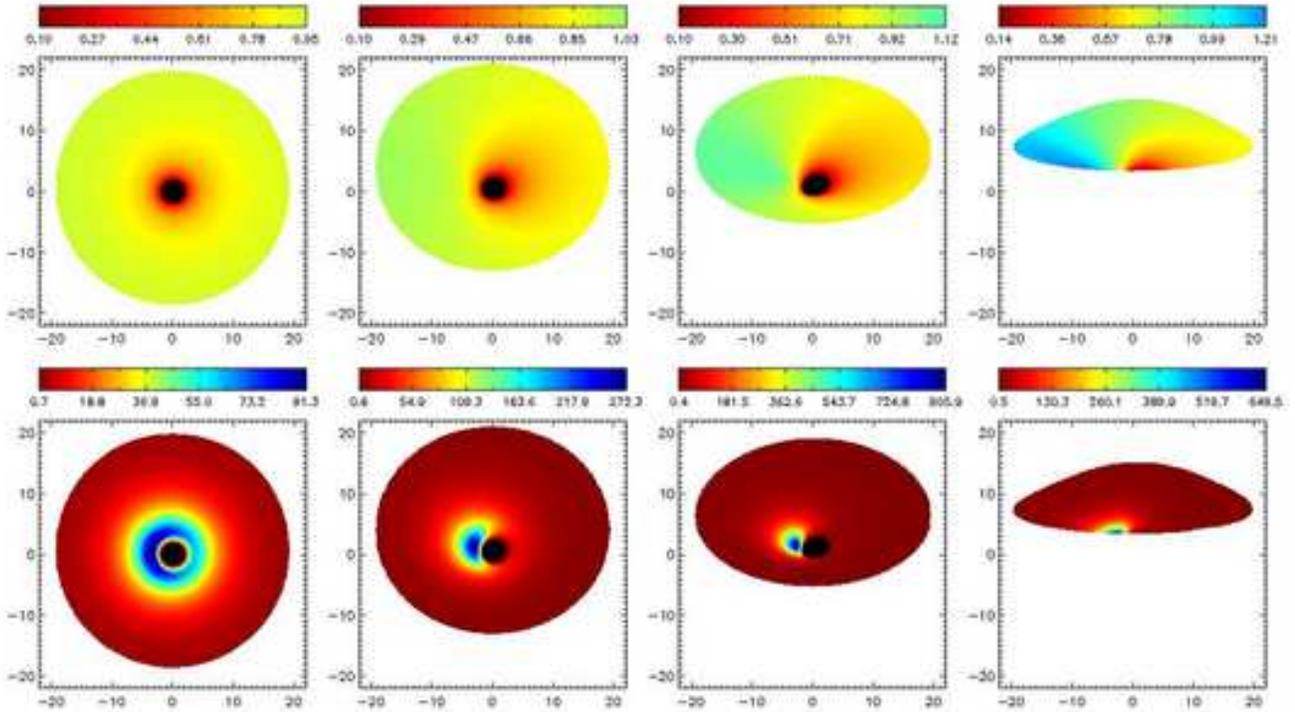}
\vspace{-1.5cm} 
\caption{As in Fig. \ref{image1} but with the angle $\vartheta_{\rm e}=70^{\circ}$.
   \label{image2}}
\end{figure*}

We present in Figs \ref{image1} and \ref{image2} the redshift and 
flux images of the accretion disc and black hole
shadows on the ($\alpha,\beta$) plane for an extreme Kerr black hole,
for $\vartheta_{\rm o}=5^{\circ}, 30^{\circ}, 55^{\circ}, 80^{\circ}$, 
and $\vartheta_{\rm e}=60^{\circ}, 70^{\circ}$. Redshift images are 
colored by the associated values of $g$ as measured by the infinity 
observer, which is defined by the scale at the top of each image. 
Flux images are colored by $10^4\varepsilon g^4$, 
again with the scale defined at the top of each image. The parameters $n$ 
and $p$ are both set to 3 and $f(\mu_{\rm e}) \propto (1+2.06 \mu_{\rm e})$. 
The images are distorted by the combined action of Doppler effects, 
gravitational redshift and light bending in the vicinity of the black 
hole. Note that at small inclination angle~($\vartheta_{\rm o}=5^{\circ}$), 
the observed radiation is all redshifted, and therefore the emission line 
profiles will have a net redshift. On the other hand, at an intermediate 
inclination angle~($\vartheta_{\rm o}=30^{\circ},\,55^{\circ}$), the 
innermost part of the disc is notably redshifted, whereas the observed 
radiation from the approaching side is remarkably enhanced by the 
Doppler boost. Moreover, the light ray emitted by the far side 
of the disc is bent by the gravity of the black hole, resulting in  
the vertical asymmetry of the image, as if it were bent toward the
observer. Note also that the self-shadowing effect is remarkable
at a high inclination angle ($\vartheta_{\rm o}=80^{\circ}$), and therefore
the black hole shadow in this case does not appear at all. The shape, size 
and position of the black hole shadows are also affected by the 
self-shadowing, which is different from those of SSD \citep{tak04}.

\section{Summary}
\label{sum}

We have developed a computer code both to calculate the line
profiles of a relativistic thick accretion disc around a black hole
and to generate the images of accretion discs. The code includes
all relativistic effects. It also includes the effect of
self-shadowing of the disc, {\rm i.e.}~the outer disc blocks the
emission from the inner region. The code can handle any value
of the black hole spin, the different viewing angle, the disc inner
radius ($r_{\rm in} \geq r_{\rm ms}$), and the disc
thickness~($\delta \leq \pi/4$). It also allows the user to choose
one of the three types of angular emissivity laws: isotropic
emission, limb-darkening or limb brightening laws.

We show that the separation and the relative height between the blue
and red peaks of the line profiles diminish as the thickness of the
disc increases because of the sub-Keplerian motion. The angular
emissivity form has also a significant influence on the line profile. 
The results of one peak line profile present in intermediate viewing 
angle in our model is different from those in SSD for low viewing angle. 
To see the self-shadowing effect more clearly, images of the disc
and the black hole shadows are also presented in this paper. The
self-shadowing effect is very important for high inclination angle. 
Future X-ray observations of high state accreting systems such 
as narrow line Seyfert 1 galaxies will be important to test whether 
the disc in these systems are indeed thick. 

Here we just present a simple disc model with a conical surface aimed
at getting an insight into the effects of geometric and dynamic
influence on the line profiles and disc images. For a non-equatorial
disc, we consider the self-shadowing and sub-Keplerian effects on
them, as well as the contribution of the light with carter constant 
${\cal Q} < 0 $ which different from those for equatorial disc. 
For simplicity, in this paper we neglected the influence of radial
drift of a flow on the line profile. Other limitations include the 
thickness of the disc may vary with radius, it probably also has a 
substantial warp, and the effects of photoionization of the surface 
layers of the accretion disc on the emission lines are not taken into 
account. X-ray reflection by photoionized accretion discs has been 
investigated in some detail \citep{ros93,ros05,ros99,ball01}. The 
ionization parameter has clearly a large 
effect on emissions lines. Evidence for reflection by ionized accretion 
discs in NLS1 has been accumulated in the literature in recent years 
\citep[see e.g.][]{bal01,bol02,bol03,fab04,gal07}. Furthermore, the radial 
drift of the flow for a sub-Keplerian disc may has significantly influence 
on the line profiles. A more realistic disc model should take into account 
both the sub-Keplerian and radial velocity effects on the line profiles. 
This effect will be investigated in the near future.

\section{acknowledgments}

We acknowledge Roland Speith for sharing his computer code. We would like to
thank the editor for useful suggestions which improve and clarify our paper. 
We would also like to thank the anonymous referee for his/her helpful comments.

\onecolumn
\begin{appendix}
\section{Integration of Photon Orbits in a Kerr Spacetime}
\label{integral}

The general orbit of a photon (indeed for any particle) in a Kerr
spacetime is described by three constants of motion \citep{car68}:
the energy-at-infinity $E$, the angular momentum about the axis of
the black hole $L_z$, and carter's constant ${\cal Q}$ . Let us write
$L_z = \lambda E$ and ${\cal Q} = Q E^2$. Then, the equations
governing the orbital trajectories are separable. Since the system
is stationary and axisymmetric, only the motions in the $r$- and
$\vartheta$-directions are required in the calculation of the
radiation spectrum from the disc. The motion in the $r-\vartheta$
plane is governed by \citep{bar72,cha83}
\begin{eqnarray}
    \int_{r_e}^r \frac{dr}{\sqrt{R(r)}} = \pm
        \int_{\vartheta_e}^\vartheta \frac{d\vartheta}{\sqrt{\Theta(
       \vartheta)}} \;,
    \label{r-th}
\end{eqnarray}
where
\begin{eqnarray}
    R(r) &=& r^4 + \left(a^2-\lambda^2-Q\right)r^2 +2\left[Q+(\lambda-a)^2
        \right]r - a^2 Q \;, \label{rr} \\[2mm]
    \Theta(\vartheta) &=& Q + a^2 \cos^2\vartheta - \lambda^2 \cot^2\vartheta\;,
        \label{thth}
\end{eqnarray}
and $r_e$ and $\vartheta_e$ are the starting values of $r$ and
$\vartheta$.

Define $\mu = \cos\vartheta$, then equation~(\ref{r-th}) becomes
\begin{eqnarray}
    \int_{r_e}^r \frac{dr}{\sqrt{R(r)}} = \pm
        \int_{\mu_e}^\mu \frac{d\mu}{\sqrt{\Theta_\mu(\mu)}} \;,
    \label{r-mu}
\end{eqnarray}
where $\mu_e = \cos\vartheta_e$ and
\begin{eqnarray}
    \Theta_\mu(\mu) = Q + (a^2-\lambda^2-Q)\mu^2 - a^2\mu^4 =
        a^2\left(\mu_-^2+\mu^2\right)\left(\mu_+^2-\mu^2\right)\;,
        \label{thmu}
\end{eqnarray}
and $\mu_\pm^2$ are defined by
\begin{eqnarray}
    \mu_\pm^2 = \frac{1}{2a^2}\left\{\left[\left(\lambda^2+Q-a^2\right)^2
        +4a^2Q\right]^{1/2}\mp\left(\lambda^2+Q-a^2\right)\right\} \;.
    \label{mupm}
\end{eqnarray}
For $Q>0$, both $\mu_+^2$ and $\mu_-^2$ are non-negative. when $Q<0$,
$\mu_-^2$ is less then zero. Note, $\mu_+^2 \mu_-^2 = Q/a^2$.

For a photon emitted by the disc, the integral over $\mu$ can be
worked out with the inverse Jacobian elliptic integral
\begin{eqnarray}
    \int_\mu^{\mu_+}\frac{d\mu}{\sqrt{\Theta_\mu}}&=&
        \int_\mu^{\mu_+}\frac{d\mu}{\sqrt{a^2\left(\mu_-^2+\mu^2\right)
       \left(\mu_+^2-\mu^2\right)}}\nonumber \\[4mm]
       &=& \left\{\begin{array}{ll}\DF{1}{\sqrt{a^2\left(\mu_+^2+
        \mu_-^2\right)}}\,\sn^{-1} \left(\left.\sqrt{\DF{\mu_+^2-\mu^2}{\mu_+^2}}\right|
        \DF{\mu_+^2}{\mu_+^2+\mu_-^2}\right) \;,& (\mu_-^2>0) \\[6mm]
       \DF{1}{a\,\mu_+}\,\sn^{-1} \left(\left.\sqrt{\DF{\mu_+^2-\mu^2}{\mu_+^2+\mu_-^2}}\right|
        \DF{\mu_+^2+\mu_-^2}{\mu_+^2}\right) \;,& (\mu_-^2<0)\\[2mm]
       \end{array}\right.
    \label{mu_int}
\end{eqnarray}
where\ $0\leq \mu < \mu_+$\ for\ $\mu_-^2>0$\ and\
$\sqrt{-\mu_-^2}\leq \mu < \mu_+$\ for\ $\mu_-^2<0$.

The integral over $r$ can also be worked out with inverse Jacobian
elliptic integrals. To do so, we need to find out the four roots of
$R(r)=0$. $R(r)=0$ may has four real roots; two real roots and two
complex roots or four complex roots. We consider the three cases
separately:

{\em Case A. $R(r)=0$ has four real roots~~} Let us denote the four
roots by $r_1$, $r_2$, $r_3$ and $r_4$ in the descending order.

The integral over $r$ can be worked out by the following integration
\begin{eqnarray}
    \int_{r_1}^r \frac{dr}{\sqrt{R}} &=& \int_{r_1}^r \frac{dr}{\sqrt{
            (r-r_1)(r-r_2)(r-r_3)(r-r_4)}} \nonumber\\[2mm]
        &=& \frac{2}{\sqrt{(r_1-r_3)(r_2-r_4)}}\, \sn^{-1} \left[\left.
           \sqrt{\frac{(r_2-r_4)(r-r_1)}{(r_1-r_4)(r-r_2)}}\right| m_4
           \right] \;,
    \label{r4_int1}
\end{eqnarray}
where
\begin{eqnarray}
  \hspace{-0cm}  m_4 = \frac{(r_1-r_4)(r_2-r_3)}{(r_1-r_3)(r_2-r_4)} \;.
    \label{m4}
\end{eqnarray}
When $r_1 = r_2$, the integral over $r$ can be expressed in terms of
a logarithm ,which is of no practical interest for this paper.

{\em Case B. $R(r)=0$ has two complex roots and two real roots~~}
Let us assume that $r_1$ and $r_2$ are complex, $r_3$ and $r_4$ are
real and $r_3 > r_4\,$. The physically allowed region for photons is
given by $r > r_3\,$. If we write $r_1$ and $r_2$ in the form
$r_1=u+iv$ and $r_1=u-iv\,$, the integral over $r$ can be worked out
with the following integration
\begin{eqnarray}
    \int_{r_3}^r \frac{dr}{\sqrt{R}} &=& \int_{r_3}^r \frac{dr}{\sqrt{
           [(r-u)^2 +v^2](r-r_3)(r-r_4)}} \nonumber\\[6mm]
        &=& \left\{\begin{array}{ll}\DF{1}{\sqrt{pq}}\,\sn^{-1} \left[\left.
       \DF{2\sqrt{pq(r -r_3)(r -r_4)}}{(p+q)r -r_3 q -r_4 p}\right|m_2\right]\;,&(r<r_c) \\[6mm]
       \DF{1}{\sqrt{pq}}\left(2K(m_2) -\sn^{-1} \left[\left.\DF{2\sqrt{pq(r -r_3)(r -r_4)}}
       {(p+q)r -r_3 q -r_4 p}\right|m_2\right]\right)\;,& (r>r_c)\\[2mm]
       \end{array}\right.
    \label{r2_int}
\end{eqnarray}
where
\begin{eqnarray}
    p^2 &=& (r_3-u)^2 + v^2 \;, \hspace{1.cm} q^2 = (r_4-u)^2 + v^2  \;,\nonumber\\[2mm]
    r_c &=& \frac{r_3q -r_4p}{q-p}\;, \hspace{1.8cm}m_2 = \frac{(p+q)^2-(r_3-r_4)^2}{4pq}\nonumber \;,
\end{eqnarray}
and
\begin{eqnarray}
  \hspace{-0cm}  K(m_2) = \sn^{-1} \left(\left.1\right|m_2\right).
\end{eqnarray}
  $K(m_2)$ is the complete elliptic integral of the first kind. When
$r = \infty$, equation~(\ref{r2_int}) is  equivalent to
equation~(28) of \citet{cad98}. It can be directly verified that
\begin{eqnarray}
    K(m_2) -\sn^{-1}\left(\left.\DF{2\sqrt{pq}}{p+q}\right|m_2\right) =
     \sn^{-1} \left(\left.\sqrt{1-1/\lambda_1}\right|m_2\right).
\end{eqnarray}

 {\em Case C: $R=0$ has four complex roots~~} Let us denote the
four roots by $r_1$, $r_2$, $r_3$ and $r_4$. $r_1 = r_2^\star$, $r_3
= r_4^\star$, where $\star$ stands for complex conjugate. If we set
$B=|r_1 +r_2|$, $C=r_1r_2$, $D=r_3r_4$, which leads to
$R(r)=(r^2-Br+C)(r^2+Br+D)$. In this case, in order to express the
quartic in terms of $r^2$, we make the substitution
\[   \hspace{-0cm}   s=\frac{r-\lambda_2}{r-\lambda_1}, \]
where $\lambda_1,\lambda_2 (\lambda_1<\lambda_2)$ is the two real
roots of equation
\[   \hspace{-0cm}   2B\lambda^2 +2(D-C)\lambda - B(D+C) =0. \]
This yields
\begin{eqnarray}
      R(r)=\frac{(p_1s^2+q_1)(p_2s^2+q_2)}{(s-1)^4}=\frac{p_1p_2(s^2+p^2)(s^2+q^2)}{(s-1)^4},
      \label{rs_int}
\end{eqnarray}
where
\[\begin{array}{ll}
   p_1=\lambda_1^2 -B\lambda_1 +C, &q_1=\lambda_2^2 -B\lambda_2 +C,\\
   p_2=\lambda_1^2 +B\lambda_1 +D, &q_2=\lambda_2^2 +B\lambda_2 +D,\\
   p^2 = {\rm max}\{q_1/p_1, q_2/p_2\},  &q^2 = {\rm min}\{q_1/p_1,
   q_2/p_2\}.
   \end{array}\]
From equation~(\ref{rs_int}) the integral over $r$ can be calculated
by means of the substitution $s=\frac{r-\lambda_2}{r-\lambda_1}$
through the following formula
\begin{eqnarray}
    \int_{r_e}^\infty \frac{dr}{\sqrt{R}} &=& \int_{r_e}^\infty \frac{dr}{\sqrt{
           (r^2 -Br+C)(r^2 +Br+D)}} \nonumber\\[6mm]
        &=& \left\{\begin{array}{ll}\DF{\lambda_2-\lambda_1}{\sqrt{p^2p_1p_2}}\,\Bigg(\sn^{-1} \left(\left.
       \sqrt{\DF{1}{1+q^2}}\right|m_c\right)-\sn^{-1}\left(\left.
       \sqrt{\DF{s^2}{s^2+q^2}}\right|m_c\right) \Bigg
       )\;,(r_e>\lambda_2) \\[6mm]
       \DF{\lambda_2-\lambda_1}{\sqrt{p^2p_1p_2}}\,\Bigg(\sn^{-1} \left(\left.
       \sqrt{\DF{1}{1+q^2}}\right|m_c\right)+\sn^{-1} \left(\left.
       \sqrt{\DF{s^2}{s^2+q^2}}\right|m_c\right)\Bigg)\;,(r_e<\lambda_2) \\[2mm]
       \end{array}\right.
    \label{rc_int}
\end{eqnarray}
where
\begin{eqnarray}
    m_c = \frac{p^2 -q^2}{p^2}.
\end{eqnarray}
\end{appendix}
\label{lastpage}
\end{document}